\documentclass[amsmath,amssymb,aps,prl,reprint,superscriptaddress]{revtex4-2}

\usepackage[colorlinks=true,allcolors=blue,breaklinks=true]{hyperref} % in-document references
\usepackage{xurl}
\usepackage{orcidlink}
\usepackage{bm} % Bold math
\usepackage[normalem]{ulem}
\newcommand{\one}{I}

\usepackage[ISO]{diffcoeff}[=v4]% nice derivatives
\diffdef{}{long-var-wrap = dv}

\newcommand{\fdv}[3][1]{\diff.delta.[#1]{#2}{#3}}

\newcommand{\dd}{\mathrm{d}}
\newcommand{\E}[1]{\left \langle #1 \right \rangle}
\newcommand{\Oh}{\mathcal{O}}
\newcommand{\T}{\mathcal{T}}

\newcommand{\D}{\mathcal{D}}
\newcommand{\I}{\mathrm{Im}}

\newcommand{\M}{\mathcal{M}}
\newcommand{\m}{\mathfrak{m}}
\newcommand{\Ge}{\mathcal{G}}
\newcommand{\Te}{\mathfrak{T}}
\newcommand{\TI}{{\mathcal{T}_\mathrm{I}}}
\newcommand{\TII}{{\mathcal{T}_\mathrm{II}}}
\newcommand{\TPI}{{\mathcal{T}_{P\mathrm{I}}}}
\newcommand{\TP}{{\mathcal{T}_{P}}}

\usepackage{feynmp-auto}
\begin{document}
\begin{fmffile}{feyn}

\title{Fluctuation Dissipation Relations for Active Field Theories} 

\author{Martin Kj{\o}llesdal Johnsrud\,\orcidlink{0000-0001-8460-7149}}
\affiliation{Max Planck Institute for Dynamics and Self-Organization (MPI-DS), D-37077 G\"ottingen, Germany}

\author{Ramin Golestanian\,\orcidlink{0000-0002-3149-4002}}
\affiliation{Max Planck Institute for Dynamics and Self-Organization (MPI-DS), D-37077 G\"ottingen, Germany}
\affiliation{Rudolf Peierls Centre for Theoretical Physics, University of Oxford, Oxford OX1 3PU, United Kingdom}

\date{\today}

\begin{abstract}
Breakdown of time-reversal symmetry is a defining property of non-equilibrium systems, such as active matter, which is composed of units that consume energy. We employ a formalism that allows us to derive a class of identities associated with the time-reversal transformation in non-equilibrium field theories, in the spirit of Ward-Takahashi identities. 
We present a generalization of the fluctuation-dissipation theorem valid for active systems as a particular realization of such an identity, and consider its implications and applications for a range of active field theories.
The field theoretical toolbox developed here helps to quantify the degree of non-equilibrium activity of complex systems exhibiting collective behavior.
\end{abstract}

\maketitle

\textit{Introduction---}Symmetry with respect to time-reversal is a necessary requirement for a system to reach thermodynamic equilibrium. It gives rise to strong constraints on the dynamics, such as the fluctuation-dissipation theorem (FDT) that connects the stochastic fluctuations of a thermal system at equilibrium and its response to external perturbations, which can be implemented as a supersymmetry in the field-theoretical context \cite{zinn-justinQuantumFieldTheory1989}.
This powerful relation can however no longer be relied upon in non-equilibrium conditions corresponding to externally driven systems
\cite{LetPeliti1997,LetDean1997,Golestanian2002,crisantiViolationFluctuationDissipation2003,haradaEqualityConnectingEnergy2005,haradaEnergyDissipationViolation2006,speckRestoringFluctuationdissipationTheorem2006,Maes2009,Parrondo2009,Harada2009,Maes2020}, as well as active matter \cite{gompper2020}, which comprises synthetic and biological systems that can produce mechanical and chemical activity \cite{julicherModelingMolecularMotors1997,ramaswamyMechanicsStatisticsActive2010,mugnai2020theoretical,borsley2022chemical,Pumm2022,Shi2022,Shi2023,Golestanian2019phoretic}. A question of particular importance is how to quantify the degree of non-equilibrium activity in such active systems \cite{Fodor2016,battle2016broken,nardini2017entropy,pietzonka2017entropy,shankar2018hidden,klappNonreciprocalInteractionLiving2023}. 
Possible scenarios to achieve this goal can build on strategies exploited in existing non-equilibrium generalization of the FDT \cite{haradaEqualityConnectingEnergy2005,haradaEnergyDissipationViolation2006,Harada2009,Gingrich2016,nardini2017entropy,yolcuGeneralFluctuationResponse2017,Golestanian2025,maesCanonicalStructureDynamical2008}, which we will hereby generically label as fluctuation dissipation relations (FDRs).

Phenomenological descriptions of active matter systems have provided us with a rich variety of active field theories, including the Toner-Tu equations for polar flocks \cite{tonerLongRangeOrderTwoDimensional1995}, active nematics \cite{aditisimhaHydrodynamicFluctuationsInstabilities2002}, field theories describing collective chemotaxis and growth \cite{gelimsonCollectiveDynamicsDividing2015,mahdisoltaniNonequilibriumPolarityinducedChemotaxis2021,BenAlZinati2021}, the active model B(+) \cite{wittkowskiScalarF4Field2014,tjhungClusterPhasesBubbly2018}, active model H \cite{tiribocchiActiveModelScalar2015}, active field theories driven by persistent noise~\cite{paoluzziScalingEntropyProduction2022,paoluzziNoiseInducedPhaseSeparation2024}, and the Non-Reciprocal Cahn-Hilliard (NRCH) model \cite{sahaScalarActiveMixtures2020,youNonreciprocityGenericRoute2020}. These active field theories exhibit a plethora of features out of reach for equilibrium and externally driven systems, such as the existence of true long-range polar order associated with a continuous symmetry breaking in two dimensions \cite{tonerLongRangeOrderTwoDimensional1995} and emergent long-range polar order in scalar systems even in two dimensions \cite{pisegnaEmergentPolarOrder2024}, which are in violation of the expectations from the Mermin-Wagner theorem, among others. Active field theories can moreover describe counter-intuitive phenomena such as condensation without attractive interactions \cite{catesMotilityInducedPhaseSeparation2015,RG2019}, and open new biological frontiers by providing spatiotemporal control capabilities in complex metabolically active mixtures \cite{parkavousiEnhancedStabilityChaotic2025}. Therefore, active field theories provide a promising and rich setting for the development of techniques for quantitative studies of non-equilibrium behavior at the coarse-grained level.

In this Letter, we consider a general class of active field theories described by $\varphi_a(\bm x, t)$, representing the physical fields of the system, and $\tilde \varphi_a(\bm x, t)$, representing the corresponding response fields, in the context of the Martin-Siggia-Rose Janssen-De Dominicis response-field formalism~\cite{zinn-justinQuantumFieldTheory1989}. Using the transformation properties of the field theories under time- and parity-reversal operation, we derive exact identities that connect the correlation and response functions of the system to expectation values of quantities that involve the stochastic entropy production of the field theory $S[\varphi]$, by following prescriptions similar to those used in deriving Ward-Takahashi identities associated with continuous symmetries.

To demonstrate applications of the Ward-like identities, we focus on two-point correlations, and derive relations concerning the correlation functions, defined in real space as $C_{ab}(\bm x-\bm x',t-t')\equiv\E{\varphi_a(\bm x, t) \varphi_b(\bm x', t')}$, as well as the the linear response of the system to a perturbation field $h(\bm x, t)$ quantified by the susceptibility $\chi_{ab}$, which is defined in Fourier space via $\chi_{ab}(\bm q, \omega) \equiv \fdv{\E{\varphi_a(\bm q, \omega)}}{{h_b(-\bm q, -\omega)}}$. We show that for an active system these quantities obey the following {\em exact} identities, namely
\begin{align}
    & C_{ab}(\bm x-\bm x',t-t')-C_{ba}(\bm x-\bm x',t-t')  =\nonumber \\
    & \hskip2.3cm\E{\varphi_a(\bm x, t) \varphi_b(\bm x', t') \left(e^{-S[\varphi]}-1\right) },\label{eq: C-asym}
\end{align}
in real space, as well as
\begin{align}
 & \chi_{ab}(\bm q, \omega)-\chi_{ab}(-\bm q, -\omega) - i \omega \beta \,C_{ab}(\bm q, \omega)=\nonumber \\
    & \hskip1.7cm    \Gamma \E{\varphi_a(-\bm q,-\omega) i \tilde \varphi_b(\bm q, \omega)  \left(e^{-S[\varphi]}-1\right)},\label{eq: GFDT-1}
\end{align}
and
\begin{align}
 & \chi_{ab}(\bm q, \omega)-\chi_{ba}(-\bm q, -\omega) - i \omega \beta \,C_{ab}(\bm q, \omega)=\nonumber \\
    & \hskip1.05cm    \frac{\Gamma^2}{i \omega \beta} \E{ i \tilde \varphi_a(-\bm q,-\omega) i \tilde  \varphi_b(\bm q, \omega) \left(e^{-S[\varphi]}-1\right)},\label{eq: GFDT-2}
\end{align}
in Fourier space, where $\Gamma$ is the mobility of the system and $\beta=\Gamma/D$, where $D$ is the noise strength.
We observe that for equilibrium systems, where $S = 0$, Eq. \eqref{eq: C-asym} recalls the symmetry of the correlation functions, $C_{ab}^{\rm eq}(\bm x,t)=C_{ba}^{\rm eq}(\bm x,t)$, 
whereas Eqs. \eqref{eq: GFDT-1} and \eqref{eq: GFDT-2} recover the familiar FDT, 
namely, $2 \I \chi_{ab}^{\rm eq}(\bm q, \omega) = \omega \beta C_{ab}^{\rm eq}(\bm q, \omega)$, as well as the symmetry properties of the susceptibility, $\chi_{ba}^{\rm eq}(\bm q, \omega)=\chi_{ab}^{\rm eq}(\bm q, \omega)$. 
These identities make powerful statements as they quantify the departure from equilibrium of a system due to internal driving that is a hallmark of active systems.
This is in contrast to fluctuation theorems where an external force protocol drives the system \cite{seifertStochasticThermodynamicsFluctuation2012}, such as for the related identities derived in~\cite{mallickFieldtheoreticApproachNonequilibrium2011,aronSymmetriesGeneratingFunctionals2010}.
They provide us with a fine-grained measure of deviation from equilibrium at all frequencies and wave-numbers.
The formalism introduced in this Letter is applicable to a wide range of scenarios and models.
We discuss some of the generalizations and avenues of further research, and in a companion paper~\cite{johnsrudFluctuationDissipationRelations2025} we showcase its applications in the context of the NRCH.

\textit{Deriving the identities---}Consider a model governed by a general overdamped Langevin field equation
\begin{align}
    \label{eq:  model}
    \partial_t \varphi_a(\bm x, t) 
    = \Gamma K_a[\varphi](\bm x, t) + \sqrt{2 D} \, \xi_a(\bm x, t),
\end{align}
where $\xi_a$ represent unit white noise
What follows will equally hold for conserved models by mapping $\Gamma\rightarrow -\nabla^2 \Gamma$ and $D \rightarrow -\nabla^2 D$. 
For now, we consider $\Gamma$ and $D$ to be constants, and, assume that the Einstein relation applies, namely, $\Gamma = \beta D$.

The corresponding response-field action reads
\begin{align}\label{eq: response action}
    &A[\varphi, \tilde \varphi]  = 
    \int\limits_{t,\bm x}
    \left\{
        i \tilde \varphi_a
        \Big(
            \partial_t \varphi_a - \Gamma K_a[\varphi]
        \Big)
        + \tilde \varphi_a D \tilde \varphi_a
        \right\},
\end{align}
where summation is implied over the index $a$, and a shorthand is used for integrals $\int_{t, \bm x} \equiv \int \dd t \int \dd^d {\bm x}$ over all of space and time. The expectation value of a given functional of the field configurations, $\Oh[\tilde \varphi, \varphi]$, is
\begin{align}
    \E{\Oh[\tilde  \varphi, \varphi]} = \int \D \tilde \varphi \D  \varphi \, \Oh[ \tilde \varphi, \varphi] e^{-A[ \tilde \varphi, \varphi]}.\label{eq:O-def}
\end{align}
We derive the Ward-like identity by applying the following transformation 
\begin{align}
    \label{eq: transformation}
    \T:
    \begin{cases}
        \tilde \varphi_a(\bm x, t) \rightarrow 
        \tilde \varphi_a(- \bm x, -t) 
        + 
        \frac{i}{D} \partial_t \varphi_a(- \bm x, -t),\\
        \varphi_a(\bm x, t) \rightarrow \varphi_a(- \bm x, -t),
    \end{cases}
\end{align}
which has been used in derivations of fluctuation and response relations in glassy \cite{andreanovDynamicalFieldTheory2006} and non-equilibrium systems \cite{aronSymmetriesGeneratingFunctionals2010}, and found applications in other classical and quantum non-equilibrium studies \cite{canetNonperturbativeApproachCritical2007,zelleUniversalPhenomenologyCritical2024,davietNonequilibriumCriticalityOnset2024,siebererThermodynamicEquilibriumSymmetry2015}.
This transformation is an involution, i.e. $\T^2$ is the identity transformation, as is necessary for time reversal.

We consider active forces of the general form $K_a = -\fdv{F}{\varphi_a} + W_a$, i.e., composed of a part that derives from a free energy $F$, as well as a non-conservative part $W_a$.
Importantly, the fact that $W_a$ cannot be written as a functional derivative breaks the $\T$-invariance of $A$. The difference in action $A$ between a forward path $(\tilde \varphi, \varphi)$ and the corresponding transformed path, $(\T \tilde \varphi, \T \varphi)$ is given by the total entropy production operator
\begin{align}
    \label{eq: entropy}
    S[\varphi] = \beta\int\limits_{t,\bm x} \partial_t \varphi_a(\bm x, t) W_a[\varphi](\bm x, t).
\end{align}
By applying $\T$ to Eq. \eqref{eq:O-def}, we obtain \cite{supp}
\begin{align}
    \label{eq: Ward identity}
    \E{\Oh[\T \tilde  \varphi, \T \varphi] } 
    = \E{\Oh[\tilde   \varphi, \varphi] e^{-S[\varphi]}},
\end{align}
which allows us to generate an arbitrary number of Ward-like identities that connect correlations and responses to the entropy production, thereby generalizing the integral fluctuation theorem that corresponds to the specific choice of $\Oh=1$.

To proceed further, we introduce a compact notation and define the vector $(\psi_1,\cdots) = (i\tilde\varphi_1, \cdots ,\varphi_1,\cdots)$ that contains both the response and physical fields. To investigate two-point functions, we choose $\Oh = \psi_i \psi_j$, and introduce the following definition 
\footnote{We denote $\delta_{\bm q+\bm q'} \equiv (2 \pi)^{d} \delta^d(\bm q + \bm q') $ and $\delta_{\omega + \omega'} \equiv 2 \pi  \delta(\omega + \omega')$ for brevity.}
\begin{align}\label{eq: definition Delta}
   \hskip-.1cm
   \E{\psi_i(\bm q, \omega) \psi_j(\bm q'\!, \omega')\left(e^{\!-S}\! - 1\right)}  
   \!\equiv\! 
   \Delta_{ij}(\bm q, \omega)   \delta_{\bm q+\bm q'} \delta_{\omega +\omega'}\!.
  \hskip-.1cm
\end{align}
The above expressions include the response propagator $G_{ab}(\bm q,\omega) \delta_{\bm q+\bm q'} \delta_{\omega+\omega'} = \E{\varphi_a(\bm q, \omega) i \tilde \varphi_b(\bm q', \omega')}$ (or the Green's tensor), which is related to susceptibility by $\chi_{ab}(\bm q, \omega) = \Gamma G_{ab}(\bm q, \omega) $. Following straightforward calculations starting from Eq. \eqref{eq: Ward identity}, we obtain four matrix identities, depending on whether the $\psi$ components are the regular fields $\varphi$ or the response fields $\tilde \varphi$, which can be written in a matrix-block form as follows \cite{supp}
\begin{align}
    \label{eq: GFDT matrix} 
    \begin{bmatrix}
        \frac{i \omega}{D}\!
        \left(\!
            G_+
            \!- G_+^\dagger
            \!- \frac{i \omega}{D} C_+
        \!\right)
        &
        G_+^\dagger
        \!- G_-^\dagger
        \!+ \!\frac{i \omega}{D} C_+^\dagger
        \\[.3cm]
        G_+
        \!- G_-
        \!- \frac{i \omega}{D} C_+ 
        & C_+ \!- C_-
    \end{bmatrix}\!
    =
    \Delta_-.
\end{align}
Here, the sign in the subscript denote the sign of the arguments, e.g., $C_{\pm, ab}(\bm q, \omega) = C_{ab}(\pm \bm q, \pm \omega)$, and similar for the Green's tensor $ G_{ab}(\bm q, \omega)$. We note that this notation naturally carries over to time-domain, as $f(-t)$ is the Fourier transform of $f(-\omega)$. In Fourier space, changing the signs of the arguments amounts to complex conjugation. The dagger denotes Hermitian conjugate, which corresponds to changing the signs of the arguments and taking the transpose, namely $G_\pm^\dagger \equiv \left(G_\pm\right)^\dagger = G_\mp^T$. Eqs.~\eqref{eq: C-asym}, \eqref{eq: GFDT-1}, and \eqref{eq: GFDT-2} can be read from the matrix elements of Eq.~\eqref{eq: GFDT matrix}. 

\textit{Active forces---}The formulas quantifying the deviations ($\Delta$) from the equilibrium FDT, namely Eq.~\eqref{eq: GFDT matrix}, as well as their further generalizations, are the main results of this paper. To gain further insight and illustrate how the identities work in practice, we now consider some specific models and applications.
For example, non-conservative forces materialize in the action-reaction symmetry breaking terms of the Non-Reciprocal Cahn-Hilliard (NRCH) model~\cite{sahaScalarActiveMixtures2020,youNonreciprocityGenericRoute2020}, which take the form $W_a = f[\varphi] \epsilon_{ab} \varphi_b$, where $\epsilon$ is the anti-symmetric Levi-Civita tensor and $f$ a functional of $\varphi$.
These forces are divergence free (in the space of fields) and cannot be written as gradients of scalar potentials, resulting in non-zero steady-state entropy production [Eq.~\eqref{eq: entropy}].

With standard methods from stochastic field theory, we can define an \emph{entropy-consumption vertex} from $ - S$ and calculate Eq.~\eqref{eq: definition Delta} perturbatively using Feynman diagrams. We note that such a strategy allows us to eliminate the entropy consumption vertex by combining Eqs.~\eqref{eq: C-asym},~\eqref{eq: GFDT-1}, and \eqref{eq: GFDT-2}, and obtain FDRs that are valid out of equilibrium, such as
\begin{align}
    G_{+} - C_{+}C_{-}^{-1} G_{-} 
    = \frac{i\omega}{D} C_{+},\label{eq:NRCH}
\end{align}
and $
\frac{i\omega}{D} ( G_+ - G_+^\dagger - \frac{i\omega}{D} C_+ ) = G_+^\dagger C^{-1}_-(C_+ - C_-)C^{-1}_-G_+
$. The details of the derivation are presented in a companion paper~\cite{johnsrudFluctuationDissipationRelations2025}.

Models with a single scalar field cannot exhibit non-reciprocal interactions, but the gradient structure may break time-reversal symmetry. 
Examples of such field theories are Kardar-Parisi-Zhang (KPZ) equation \cite{kardarDynamicScalingGrowing1986} and active model B \cite{wittkowskiScalarF4Field2014}, for which the active forces are given by 
$W = \frac{1}{2} \lambda \left[\nabla \varphi(\bm x, t)\right]^2$, whereas non-local generalizations of this term appear in field theories describing collective chemotaxis and growth \cite{gelimsonCollectiveDynamicsDividing2015,mahdisoltaniNonequilibriumPolarityinducedChemotaxis2021,BenAlZinati2021}.
In the case of the KPZ, the full force term is $K = \nu \nabla^2\varphi 
    + \frac{1}{2}\lambda(\nabla\varphi)^2$, which is not conserved.
The entropy consumption vertex is \cite{supp}
\begin{align}
    \parbox{12mm}{
    % \begin{fmfgraph*}(12,8)
    %     \setval
    %     \fmfleft{i}
    %     \fmfright{o1,o2}
    %     \fmf{fermion}{i,m}
    %     \fmf{plain}{m,o1}
    %     \fmf{plain}{m,o2}
    %     \fmfdot{m}
    % \end{fmfgraph*}
    \includegraphics{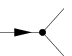} % Same as file generated above
    }
    & \equiv - i\omega_3 \beta \lambda \, \bm q_1 \cdot \bm q_2,
\end{align}
that can be used to perform a one-loop calculation, which yields
\begin{align}
    G_+ - G_- - \frac{i \omega}{D} C_+
    = \frac{D}{i\omega}G \Sigma_{22}^* G
    = - \frac{ 2 i\omega \delta D/D }{ \omega^2 + \nu^2 q^4},
\end{align}
where $\delta D \propto \int_{\bm k} \frac{1}{k^2}$ is the correction to $D$, and $\Sigma_{22}$ is the amputated version of $\Delta_{11}$~\footnote{The indices refer to whether the external legs of the diagrams are $\psi_1 = i \tilde \varphi$ or $\psi_2 = \varphi$.}.
In one dimension, the KPZ equation has a Gaussian steady state and consequently obeys an FDR (see below)
\cite{kardarDynamicScalingGrowing1986,dekerFluctuationdissipationTheoremsClassical1975}.

The map $\Gamma\rightarrow -\nabla^2 \Gamma$ and $D \rightarrow -\nabla^2 D$ yields the conserved KPZ~\cite{sunDynamicsDrivenInterfaces1989}.
This, however, breaks the Galilean symmetry that underlies the KPZ universality class~\cite{janssenCriticalExponentsRenormalization1997}, opening up the possibility of adding further symmetry-breaking terms to the equation of motion.
Active model B adds a ``mass-term'' $\frac{1}{2} r \varphi^2 $ to $F$
~\cite{janssenCriticalExponentsRenormalization1997,wittkowskiScalarF4Field2014}.
This yields a different correction $\delta D'$, which will still satisfy the FDR as $\Sigma_{22} \propto \delta D'$. Due to the conservation law, this correction turns out to be irrelevant (in the RG sense) to all orders in the perturbation theory. This approach can be extended with further terms, as is done for the active model B+~\cite{tjhungClusterPhasesBubbly2018}.

\textit{Colored noise and multiplicative noise---}%
The non-intuitive form of $\T$, Eq.~\eqref{eq: transformation}, is designed to leave the dissipative part of the response-field action, $\int_{t, \bm x} \tilde \varphi \left( D \tilde \varphi + i \partial_t \varphi \right)$, invariant.
While we have considered only its simplest form, we note that the noise strength $D$ may be generalized to have a non-local kernel, $\gamma(t, t')$, or to be multiplicative and depend on $\varphi$, in which case there are subtleties associated with the time-discretization in the path-integral formulation~\cite{depireyPathIntegralsStochastic2023}. Such generalizations have been investigated in equilibrium cases, where these terms obey the Einstein relation \cite{aronSymmetriesGeneratingFunctionals2010}. 

Multiplicative noise can be incorporated in the response-field formalism by introducing Grassmann variables to handle the resulting Jacobian, in which case $\T$ symmetry includes these new anti-commuting fields \cite{zinn-justinRenormalizationStochasticQuantization1986,aronSymmetriesGeneratingFunctionals2010}. As we will show next, the $\T$-transformation can be readily generalized to deal with field-dependent noise-strength. 

When the Einstein relation is not obeyed, there exists some freedom in how to define $\T$. We can write, $\T = \M \Te$, where we defined the operator  $\Te$ via $\Te f(\bm x, t) = f(-\bm x, -t)$, while $\M$ is a matrix defined as follows 
\begin{align}
    \M_{ij} \psi_j = 
    \begin{pmatrix}
        \delta_{a b} & - \m_{ab} \partial_t \\ 0 & \delta_{ab}
    \end{pmatrix}
    \begin{pmatrix}
        i \tilde \varphi_{b} \\ \varphi_b
    \end{pmatrix}.
\end{align}
Using this definition, we obtain the second term of the transformed response field in Eq. \eqref{eq: transformation} by setting $\m = D^{-1} \one = \beta \Gamma^{-1}\one$.
If the Einstein relation is broken, we are afforded freedom to choose one of these two transformations.
Importantly, we have $\M\Te = \Te \M^{-1}$, $\Te^2= \one$, and $\T^2 = \one$, and therefore, we observe that this transformation is an involution for any choice of the sub-matrix $\m$.
In fact, we may even choose $\m$ to be a function of $\varphi$ (the Jacobian will still be unity and $\T$ will be an involution).
This means that we can readily incorporate multiplicative noise, where $D$ depends on $\varphi$.

Within our current framework the kernel $\gamma$ can be included straightforwardly. A system driven by persistent noise characterized by a finite correlation time will feature activity. The simplest example is the active Ornstein-Uhlenbeck (AOU) model \cite{martinStatisticalMechanicsActive2021,nguyenActiveOrnsteinUhlenbeck2022}, which may be generalized to fields yielding an active scalar field theory associated with a system that features effective inertia, white noise, and a non-linear friction term $\gamma(\varphi)\partial_t \varphi$, which breaks time-reversal symmetry. This leads to a non-zero entropy-consumption operator $\sigma \propto \fdv[3]{F}{\varphi}(\partial_t \varphi)^3$~\cite{paoluzziScalingEntropyProduction2022,paoluzziNoiseInducedPhaseSeparation2024}. The appropriate form for $\T$ can be obtained within the formalism introduced here by choosing $\m = D^{-1}\gamma(\varphi)$, which enables us to derive the corresponding FDRs.

\textit{Non-symmetric mobilities and parity}---Complex environments, such as fluids with odd viscosity \cite{Avron1998}, can give rise to more general mobilities $\Gamma_{ab}$ and diffusion matrices $D_{ab}$. We observe that for the diffusion matrix $D$ only the symmetric part influences the system, as it appears in a quadratic form in the action Eq.~\eqref{eq: response action}, and therefore any anti-symmetric component has no physical effect \footnote{This assumes an infinite, translationally invariant domain. In more complex domains, an anti-symmetric part of $D$ may have observable effects, but this is beyond the scope of this Letter.}.
This is, however, not the case for the mobility matrix $\Gamma$. Consequently, we have the possibility to break the Einstein relation, and drive the system out of equilibrium via this route. For general (constant) mobilities and diffusion matrices, the two natural generalizations are 
\begin{align}
\TI : \m_{ab} &\equiv D_{ab}^{-1},&
\TII : \m_{ab} &\equiv \beta \Gamma^{-1}_{ab},
\end{align}
denoted type-I and type-II, respectively.

In addition to the effect of $\T$ on the fields, the parameters $v$ that appear in the equation of motion, for example $\Gamma_{ab}$, $D$, or $\lambda$, might have non-trivial time reversal parity, i.e., some might change sign for a physical time-reversal transformation due to their micro-physical origins. We denote this action by $Pv_i = \varepsilon_i v_i$ (no summation), where $\varepsilon_i = \pm 1$ is the parity of the parameter $v_i$.
Defining 
\begin{align}
    \T_P \equiv P \M \Te,
\end{align}
we have the full, general time reversal transformation, which is still an involution.
By generalizing the derivation of the Ward-identity [Eq.~\eqref{eq: Ward identity}] using the $\T_P$-transform, we obtain
\begin{align}
    \E{\Oh[\T \tilde \varphi, \T \varphi]}_{P} = \E{\Oh[\tilde \varphi, \varphi] e^{-S_\TP[\varphi]}}.
\end{align}
Here, $\E{\cdot}_P$ indicates averages using the action with $P$-transformed parameters, $PA[\psi;v_i] = A[\psi; \varepsilon_i v_i]$.

With a straightforward generalization of the above derivation, we obtain
\begin{widetext}
\begin{align}
    \label{eq: GFDT matrix 2}
    \begin{bmatrix}
        i \omega
        \left(
            \m G_{P,+}
            - G_{P,+}^\dagger \m ^T
            - i \omega \m C_{P,+} \m ^T
        \right) 
        &&
            G_{P,+}^\dagger
            - G_-^\dagger
            + i \omega (C_{P,+}\m^T)^\dagger
        \\[.3cm]
        G_{P,+}
        - G_-
        - i \omega C_{P,+}\m^T && C_{P,+} - C_-
    \end{bmatrix}
    =
    \Delta_{\TP-},
\end{align}
\end{widetext}
where the $P$-subscript indicates that the correlations are in an ensemble with the $P$-transformed action, 
$\Delta_{\TP, ij}(\bm q,  \omega) \delta_{\bm q + \bm q'} \delta_{\omega+\omega'} = \E{\psi_i(\bm q,  \omega) \psi_j(\bm q'\!,\omega) (e^{-S_\TP} \!-\! 1) }$,
and $S_\TP \equiv \TP A - A$.

Let us first consider the case with only even parity for the parameters. For type-I and type-II cases, we obtain
\begin{align}
    S_{\TI}
    &
    = 
    \beta 
    \int_{t, \bm x} 
    \partial_t \varphi_a 
    \left( 
        W_a
        + 
        \delta \Gamma_{ab} K_b
    \right), \\
    S_{\TII}
    &
    = 
    \beta 
    \int_{t, \bm x} 
    \partial_t \varphi_a 
    \left( 
        W_a
        + 
         \sigma_{\Gamma, ab} i \tilde \varphi_b
    \right).
\end{align}
Here, we have defined $\delta \Gamma \equiv (\beta D)^{-1}\Gamma^A$ and $\sigma_\Gamma \equiv 2 (\beta \Gamma^T)^{-1}\Gamma^A$, and assumed a generalized Einstein relation, $\Gamma^S = \beta D$, where the symmetric and anti-symmetric parts of the mobility are given as $\Gamma^S = \frac{1}{2} (\Gamma + \Gamma^T)$ and $\Gamma^A = \frac{1}{2} (\Gamma - \Gamma^T)$.

Both choices have their advantages.
If we consider time reversal in the Onsager-Machlup formalism, where the action $\Ge$ is independent of $\tilde \varphi$, and thus also independent of the choice of $\m$, we obtain $\Te \Ge[ \varphi] - \Ge[\varphi] = S_{\TI}[\varphi]$.
On the other hand, if we choose type-II transformation, we obtain FDRs of the form of Eqs.~\eqref{eq: GFDT-1} and \eqref{eq: GFDT-2}, as the physical response function is related to the response propagator in the form of $\chi = G\Gamma^T$, and \emph{not} via $D$.
In fact, in the case where $W_a = 0$, entropy production is given by only $\partial_t \varphi_a \sigma_{\Gamma} i \tilde \varphi$, and causality implies that  $\E{\tilde \varphi_a\tilde \varphi_b e^{-S_{\TII}}} = 0$.
As a consequence, we have the \emph{exact} generalization of the FDT as follows
\begin{align}\label{eq:exact-FDR}
    \chi_{ab}(\bm q, \omega)
    -
    \chi^\dagger_{ab}(\bm q, \omega)
    = i \omega \beta C_{ab}(\bm q, \omega).
\end{align}
This does not hold for the other relations, as Eqs.~\eqref{eq: C-asym} and \eqref{eq: GFDT-1} will in general have non-vanishing right-hand sides, which can thus provide an experimental protocol for detecting odd mobilities.
More specifically, if the measured susceptibility obeys Eq. \eqref{eq:exact-FDR} but \emph{not} $\chi_+ - \chi_-=i\omega\beta C_+$, then we can conclude that the out-of-equilibrium behavior stems from odd mobilities \emph{only}. Furthermore, exact relations such as these are of great use in analytical calculations, e.g. as a tool for constraining the flow under renormalization group \cite{tauberPerturbativeFieldTheoreticalRenormalization2014}.

Equation~\eqref{eq: C-asym} immediately implies that the diagonal elements of $C_{ab}$ are time reversible.
This is a consequence of the translation symmetry in space and time, i.e. $\E{\varphi_a(t) \varphi_b(t')} = \E{\varphi_a(t-(t+t')) \varphi_b(t'-(t+t')}$ (and similarly for space), which together yield $C_{+,aa} = C_{-,aa}^T$, where there is no summation implied. 
This property may, however, be lost in the case of parity transformations, as the lower right quadrant of Eq.~\eqref{eq: GFDT matrix 2} is in general non-zero also on the diagonal.

If we consider the type-I transformation, and assume that the odd mobility has odd parity, such that $P\Gamma^T = \Gamma$, the entropy production is found as
\begin{align}
\hskip-.1cm S_{\TPI}
\!=
\beta \!\!
\int\limits_{t, \bm x} \!
    \Big\{
        \partial_t \varphi_a W_a
        +\left( 2 D_{ab} i \tilde \varphi_b - \partial_t \varphi_a \right) \delta \Gamma_{ac} K_c
    \Big\}.
\end{align}
While the first term is the same as before, the $\delta \Gamma$-term is different and now has a dependence on $\tilde \varphi$.
The two terms are independently odd under $\TP$.

We now address the FDR that holds in one dimension for KPZ field theory, which arises due to a combination of three factors. Firstly, the equilibrium force, i.e. the diffusion term $\nu \nabla^2 \varphi$, is linear. Secondly, in one dimension this equilibrium term is orthogonal to the non-equilibrium drive $\frac{1}{2}\lambda (\nabla \varphi)^2$, in the sense that the integral of their product gives a boundary term.
Lastly, it is customary to assume that $P\lambda = -\lambda$ and thus consider the $\TP$ choice, which results in a ``class B FDT'' relating $C$ and $\chi$~\cite{dekerFluctuationdissipationTheoremsClassical1975}. 
This differs from the relations discussed in this Letter.
As a consequence, the Onsager-Machlup entropy $S_\mathrm{OM} = \TP \Ge - \Ge$ vanishes, since the velocity$\times$force term $\frac{1}{2} \lambda \partial_t\varphi (\nabla \varphi)^2$ is even under $\TP$.
This is, however, in many instances not physical, as the KPZ is an effective theory for non-equilibrium models such as non-reciprocal active matter~\cite{pisegnaEmergentPolarOrder2024} or driven quantum systems~\cite{siebererDynamicalCriticalPhenomena2013}, which dissipate energy in steady state. Note that the class B FDT is still valid. 
In the response-field formalism presented here, the entropy is the integral of $(2 D i \tilde \varphi - \partial_t \varphi) \frac{1}{2} \lambda (\nabla\varphi)^2$, which is odd under $\TP$.
Our formalism thus quantifies the out-of-equilibrium nature of fluctuations in a seemingly equilibrium steady-state.

\textit{Discussion---}%
Quantifying entropy production in field theories comes with inherent subtleties associated with UV-divergences (in the continuum limit) not found in the underlying microscopic theory \cite{pruessnerFieldTheoriesActive2022}.
Depending on the regularization scheme used to control the effective field theory, one will have access to a coarse-grained picture of dissipation, and not all of it. Indeed, an understanding of the microscopic physics is necessary to obtain the physical dissipation. For example, the non-equilibrium character of active matter systems such as microswimmers should be quantified via the amount of hydrodynamic dissipation, which is inherently related to the faithful implementation of momentum conservation, from the nano-scale \cite{RG2008,Mike3SS2024} to the micro-scale \cite{Babak2021,daddi2023minimum,bebon2024thermodynamics}. We observe furthermore that by exploiting the existing freedoms in defining the time-reversal transformation, such as the parity of the parameters, one obtained generalizations of entropy that might not directly relate to the physical entropy of the systems. 
This notion has also been investigated in the context of the stochastic thermodynamics of single-particles, where the choice of time-reversed ``conjugate dynamics'' can give rise to different results for the entropy production \cite{seifertStochasticThermodynamicsFluctuation2012}.
We note, however, that even an unphysical choice of $\T$ can yield useful results, as the derived identities between observables will be valid regardless of the nature of $S$. 

An earlier generalization of the FDT is the Harada-Sasa relation~\cite{haradaEqualityConnectingEnergy2005,haradaEnergyDissipationViolation2006}, which relates total entropy $\E{S}$ to the functional trace (integration and summation over the indices) of the deviation from the equilibrium FDT, $\Delta$, as defined in Eq.~\eqref{eq: definition Delta}.
This has been applied to the field theories of active model B and H~\cite{nardini2017entropy}, and may be generalized to models in the form of Eq.~\eqref{eq:  model}; see Ref. \cite{supp} for details.
Taking the functional trace of Eqs.~\eqref{eq: C-asym}-\eqref{eq: GFDT-2} gives similar ``sum-rules''~\cite{supp}.
The approach laid out here can thus complement the Harada-Sasa relation, giving the relation
\begin{align}
    \int_{\bm q, \omega} \!\!
    i \omega \Delta^*_{ a\tilde a}(\bm q, \omega)
    = 
    \int_{\bm q, \omega} \!\! 
    D \Delta^*_{\tilde a\tilde a}(\bm q, \omega)
    = 
    - \frac{\E{S}}{TL^d},
\end{align}
where $L^d$ is the volume and $T$ the total time.
Here the tilde on the index, such as $\Delta_{\tilde a \tilde a}$, is a shorthand indicating that it ranges only over the response fields $\tilde \varphi_a$, while indices without tilde range over the physical fields $\varphi_a$. 
See also~\cite{supp,johnsrudFluctuationDissipationRelations2025}.

Although $\E{S}$ is a physically relevant quantity, in field theories it is strongly divergent in both the continuum limit and the thermodynamic limit~\cite{garcia-millanRunandtumbleMotionHarmonic2021}.
Spectrally resolved quantities such as $\chi(\bm q, \omega)$ and $C(\bm q, \omega)$ are, in contrast, finite and easily accessible, both experimentally and theoretically.
The spectral identities for $\Delta_{ij}$, as well as other relations introduced in this Letter, give access to more fine-grained and well-defined measures of the departure from equilibrium. The framework developed in this Letter and its companion paper, Ref.~\cite{johnsrudFluctuationDissipationRelations2025}, provides us with tools to derive exact or approximate results, and to more generally investigate the phenomenology arising from activity. We would also like to highlight that dealing with the stochastic entropy operator can present technical challenges for field theories, and this is why we regard Eqs. \eqref{eq:NRCH} and \eqref{eq:exact-FDR} as presenting particularly important opportunities to probe FDRs without having to deal with such subtlety. Finally, we note that the framework can be extended to implement the weak noise limit, which has been recently employed to obtain analytical results on time-reversal symmetry breaking and entropy production in active field theories \cite{alstonIrreversibilityNonreciprocalSymmetryBreaking2023,suchanekIrreversibleMesoscaleFluctuations2023,suchanekEntropyProductionNonreciprocal2023}, by appropriately extending the diagrammatic expansion developed in Ref. \cite{johnsrudFluctuationDissipationRelations2025}.

In summary, we have introduced a framework to quantify the non-equilibrium character of field theories, by for example probing deviations from the FDT, and applied it to examples from the literature of active field theories. We have shown that systems with odd mobility can obey a restricted FDR for some components while others are broken. We have also exploited the freedoms that exist in defining time-reversal transformations towards the derivation of alternative classes of FDRs.
In the companion paper~\cite{johnsrudFluctuationDissipationRelations2025}, we develop a diagrammatic formalism for perturbative field theoretical verification of the FDRs.
Our framework can be readily generalized in a variety of ways to serve other types of applications in stochastic thermodynamics of fields, e.g. by putting bounds on quantities such as the anti-symmetric part of the correlation function in active field theories \cite{Ohga2023}.

\begin{acknowledgements}
MKJ would like to thank Luca Cocconi and Michalis Chatzittofi for stimulating discussion.
We acknowledge support from the Max Planck School Matter to Life and the MaxSynBio Consortium which are jointly funded by the Federal Ministry of Education and Research (BMBF) of Germany and the Max Planck Society.
\end{acknowledgements}

\end{fmffile}
\bibliography{ref,ramin_ref}
\end{document}

\title{Fluctuation Dissipation Relations for Active Field Theories \\ {\it Supplemental Material}} 

\author{Martin Kj{\o}llesdal Johnsrud}
\affiliation{Max Planck Institute for Dynamics and Self-Organization (MPI-DS), D-37077 G\"ottingen, Germany}

\author{Ramin Golestanian}
\affiliation{Max Planck Institute for Dynamics and Self-Organization (MPI-DS), D-37077 G\"ottingen, Germany}
\affiliation{Rudolf Peierls Centre for Theoretical Physics, University of Oxford, Oxford OX1 3PU, United Kingdom}

\date{\today}
\maketitle

\tableofcontents

\setcounter{equation}{0}
\setcounter{figure}{0}
\renewcommand{\theequation}{S\arabic{equation}}
\renewcommand{\thefigure}{S\arabic{figure}}

\section{Derivation of Eq. (9): Ward-identity }

We may decompose the action $A$ into the dissipative part $A_1$ and the deterministic part $A_2$, as follows
\begin{align}
    A_1[\varphi, \tilde \varphi] 
    & = \int\limits_{t,\bm x}  
    \tilde \varphi_a(\bm x, t)
    \left[i\partial_t \varphi_a(\bm x, t) + D \tilde \varphi_a(\bm x, t)\right], 
    &
    A_2[\varphi, \tilde \varphi] 
    &= -\int\limits_{t,\bm x}  
    i \tilde \varphi_a(\bm x, t) \Gamma K_a[\varphi](\bm x, t).
\end{align}
Applying $\T$ on the dissipative part yields
\begin{align} \nonumber
    A_1[\T \varphi, \T \tilde \varphi]
    & = 
    \int\limits_{t,\bm x}  
    (\T \tilde \varphi_a)(\bm x, t)
    \left[
        i\partial_t (\T \varphi_a)(\bm x, t) +  (\T\tilde \varphi)_a(\bm x, t)
    \right]\\\nonumber
    & = 
    \int\limits_{t,\bm x}  
    \left[
        \tilde \varphi_a(-\bm x, -t) + i D^{-1} (\partial_t \varphi)_a(-\bm x, -t)
    \right]
    \left[
        -i(\partial_t \varphi_a)(- \bm x, -t) 
        - D \tilde \varphi_a(- \bm x, -t)
        + i  (\partial_t \varphi)_a(- \bm x, -t)
    \right]\\
    & =
    \int\limits_{t',\bm x'}
    \left[
        iD^{-1} \partial_{t'} \varphi_a(\bm x', t') + \tilde \varphi_a(\bm x', t')
    \right]
    D \tilde \varphi_a(\bm x', t')
    = A_1[\varphi, \tilde \varphi].
\end{align}
We thus observe that $A_1$ is automatically invariant under $\T$.
For the deterministic part, we assume that there is no explicit time- or space-dependence in the forces (apart from what appears through $\varphi$). Therefore, $K[\T \varphi](\bm x, t) = K[\varphi](- \bm x, -t)$, and the second part thus transforms as follows
\begin{align}\nonumber
    A_2[\T \varphi, \T \tilde \varphi]
    & =
    -i \int\limits_{t,\bm x}  
    (\T \tilde \varphi)_a(\bm x, t) \Gamma K_a[\T \varphi](\bm x, t)\\
    & =
    -\int\limits_{t,\bm x}  
    \left[
        i \tilde \varphi_a(-\bm x, - t) 
        - D^{-1}(\partial_t \varphi)_a(-\bm x, - t)
    \right]
    \Gamma K_a[\varphi](-\bm x, -t)
    =
    A_2[\varphi, \tilde \varphi] + \delta A[\varphi, \tilde \varphi],
\end{align}
which yields a difference between the original and transformed action in the form of
\begin{align}
    \delta A[\varphi, \tilde \varphi]
    =  
    \beta
    \int\limits_{t,\bm x}  
    \partial_t \varphi_a(\bm x, t)K_a[\varphi](\bm x, t).
\end{align}
Note that the variational (conservative) part of the force dos not contribute to the difference in stationary-state conditions, namely
\begin{align}
    -\beta \int\limits_{t,\bm x} \partial_t \varphi_a(\bm x, t)
    \fdv{F[\varphi](t)}{{\varphi_a(\bm x, t)}}
    = -\beta \int \dd t \,  \odv{F[\varphi](t)}{t} 
    = 0.
\end{align}
Therefore, we have
\begin{align}
    \delta A[\varphi] 
    \equiv S[\varphi]
    =
    \beta \int\limits_{t,\bm x}
    \partial_t \varphi_a(\bm x, t) \mathcal K_a[\varphi](\bm x, t),
\end{align}
which gives the total entropy production of the system. We note that the Jacobian determinant of the transformation associated with $\T$ is unity, because the Jacobian matrix is upper triangular.

Applying this transformation to an arbitrary operator $\Oh[\varphi, \tilde \varphi]$ and using $A[\varphi, \tilde \varphi] = A[\T \varphi, \T \tilde \varphi] + S[\T \varphi]$, we find
\begin{align}
    \int \D \varphi \D \tilde \varphi \,
    \Oh[\T \varphi, \T \tilde \varphi]
    e^{-A[\varphi, \tilde \varphi]}
    = \int \D (\T \varphi) \D (\T \tilde \varphi) \,
    \Oh[\T \varphi, \T \tilde \varphi]
    e^{-A[\T \varphi, \T \tilde \varphi] - S[\T \varphi]}. 
\end{align}
which yields the Ward-like identity corresponding to time-reversal translation, namely
\begin{align}
    \label{eq: Ward-like identity}
    \E{\Oh[\T \varphi, \T \tilde \varphi]}
    &
    = \E{\Oh[\varphi, \tilde \varphi] e^{-S[\varphi]}}.
\end{align}

\section{Derivation of Eq. (11): The fluctuation dissipation relations}
\label{sec: applying id}

We apply Eq.~(1) in the main text to the different combinations of second order operators. We start with $\Oh = \varphi_a(\bm x, t) i \tilde \varphi_b(\bm x', t')$, and apply $\T$ to find the following form
\begin{align}
    \Oh[\T \varphi, \T \tilde \varphi]
    = 
    \varphi_a(-\bm x, -t) i \tilde \varphi_b(-\bm x', -t')
    -
    \frac{1}{D}\, \varphi_a(-\bm x, -t) (-\partial_{t'} \varphi)_a(-\bm x', -t').
\end{align}
The Ward-like identity Eq.~\eqref{eq: Ward-like identity} explicitly reads as
\begin{align}
    \nonumber
    \E{\varphi_a(\bm x, t) i \tilde \varphi_b(\bm x',t')}
    -
    \E{\varphi_a(-\bm x, -t) i\tilde \varphi_b(-\bm x',-t')}
    & + 
    \frac{1}{D}\E{\varphi_a( -\bm x,- t) (\partial_{t'} \varphi)_b(-\bm x', -t')}\\
    & =
    \E{
        \varphi_a(\bm x, t) i \tilde \varphi_b(\bm x', t')
        \left(1 - e^{- S[\varphi]}\right)
    }.
\end{align}
We employ time- and space-translational invariance (as there is no explicit dependence on either), and always write correlations as functions of only the difference $t-t'$ and $\bm x - \bm x'$.
The left-hand side is then given as follows
\begin{align}
    G_{ab}(\bm x - \bm x', t-t') 
    -
    G_{ab}(\bm x' - \bm x, t'-t) 
    - \frac{1}{D}\, \partial_{t'} C_{ab}(\bm x' - \bm x, t'-t).
\end{align}
We rename $t', \bm x' \leftrightarrow t, \bm x$, and obtain
\begin{align}
    G_{ab}(\bm x - \bm x', t-t') - G_{ab}(\bm x' - \bm x, t'-t) 
    + \frac{1}{D}\,\partial_{t} C_{ab}(\bm x' - \bm x, t-t')
    = 
    \E{
    \varphi_a(\bm x', t') i \tilde \varphi_b(\bm x, t)
        \left(e^{- S[\varphi]} - 1\right)
    }.
\end{align}
As in the main paper, we define
\begin{align}
    \delta_{\bm q+\bm q'} \delta_{\omega +\omega'}
   &\equiv (2 \pi)^{d+1} \delta^d(\bm q + \bm q') \delta(\omega + \omega'),&
   \E{\psi_i \psi_j (e^{-S} - 1)} &\equiv \Delta_{ij} \delta_{\bm q+\bm q'} \delta_{\omega +\omega'},
\end{align}
where $(\psi_1, ... ) = (i\tilde \varphi_1, ..., \varphi_1, ...)$.
We will denote the different ``quadrants'' of $\Delta_{ij}$ by introducing a notation in which we place tilde on some of the indices specifying that the corresponding index is to be contracted with a response field, namely:
\begin{align}
    \Delta_{ij} \equiv
    \begin{bmatrix}
        \Delta_{\tilde a \tilde b} & \Delta_{\tilde a b} \\ \Delta_{a \tilde b} & \Delta_{a b}
    \end{bmatrix}.
\end{align}
Due to the translational invariance, in Fourier space all two-point correlators are proportional to delta functions on the wavevectors and frequencies
\begin{align}
    C_{ab}(\bm q, \bm q', \omega, \omega')
    =
    \int\limits_{ t, t',\bm x, \bm x'}
    e^{i( \omega t + \omega't' - \bm x\cdot \bm q - \bm x'\cdot \bm q')}
    C_{ac}(\bm x - \bm x', t-t')
    =
    C_{ab}(\bm q, \omega) \delta_{\bm q + \bm q'} \delta_{\omega  +\omega'},
\end{align}
With this, the identity can be written as
\begin{align}
    \label{eq: FDT GC}
    % \left(
    G_{ab}(\bm q, \omega) - G_{ab}(-\bm q, -\omega)
    -  \frac{i \omega}{D} C_{ac}(\bm q, \omega)
    =
    \Delta_{a\tilde b}(-\bm q, -\omega),
\end{align}

Shifting the indices and arguments to $\Oh =i \tilde \varphi_a(\bm x, t) \varphi_b(\bm x', t')$ corresponds to taking the Hermitian conjugate, so
\begin{align}
    \label{eq: FDT 2}
    G^\dagger_{ab}(\bm q, \omega) - G^\dagger_{ab}(-\bm q, -\omega) + \frac{i \omega}{D} C_{ab}^\dagger(\bm q, \omega) 
    =
    \left[
    \Delta_{a \tilde b}(-\bm q, - \omega)
    \right]^\dagger
    =
    \Delta_{\tilde b a}(\bm q, \omega)
    =
    \Delta_{\tilde a b}(- \bm q, -\omega).
\end{align}
If we use $\Oh = \varphi_a(\bm x, t) \varphi_b(\bm x', t')$, we obtain the following identity
\begin{align}
    C_{ab}(\bm q, \omega)
    - 
    C_{ab}(-\bm q, -\omega)
    = 
    - \Delta_{ab}(\bm q, \omega)
    =
    \Delta_{ab}(- \bm q, -\omega)
    .
\end{align}
We note that $C^\dagger = C$, and thus $C_{-} = C_+^T$. This is therefore also a formula for the transpose of $C$. 
Finally, using $\Oh = i \tilde \varphi_a(\bm x, t) i \tilde \varphi_b(\bm x', t')$, we obtain the following identity
\begin{align}
    \frac{i \omega}{D}
    \left[
        G_{ab}(\bm q, \omega)
        - G^\dagger_{ab}(\bm q, \omega)
        - \frac{i \omega}{D} C_{ab}(\bm q, \omega)
    \right]
    = \Delta_{\tilde a \tilde b}(-\bm q, -\omega).\label{eq:gen-2-FDT}
\end{align}
Here, we have used the fact that $\E{\tilde \varphi \tilde \varphi} = 0$, a consequence of the casual structure of the model.
We see that Eq. \eqref{eq:gen-2-FDT} is also a generalization of the FDT, as the anti-Hermitian part of a matrix, $G - G^\dagger$, and its imaginary part, $G_+ - G_-$, both reduce to the imaginary part in the case of a scalar.
All the different identities can be summarized as follows
\begin{align}
    \label{eq: GFDT matrix}
    \begin{bmatrix}
        \frac{i \omega}{D}
        \left(
            G_+
            - G_+^\dagger
            - \frac{i \omega}{D} C_+
        \right) 
        &
            G_+^\dagger
            - G_-^\dagger
            + \frac{i \omega}{D} C_+^\dagger
        \\[3mm]
        G_+
        - G_-
        - \frac{i \omega}{D} C_+ & C_+ - C_-
    \end{bmatrix}
    =
    \Delta_-.
\end{align}

\section{Derivation of Eq. (14): Kardar-Parisi-Zhang field theory}
\begin{fmffile}{supp/feyn-supp}

The KPZ equation takes the form
\begin{align}
    \partial_t \varphi = 
    \nu \nabla^2\varphi 
    + \frac{\lambda}{2}(\nabla\varphi)^2 + \sqrt{2D} \,\xi.
\end{align}
The non-linear $\lambda$-term cannot be written as a derivative, and the entropy production is as follows
\begin{align}
    S[\varphi] = \frac{1}{2}\lambda \beta \int_{\bm x, t} \partial_t \varphi (\nabla \varphi)^2.
\end{align}
The diagrammatic expansion of the theory can be constructed using the following elements
\begin{align}
    \parbox{15mm}{
        \begin{fmfgraph*}(15,10)
            \setval
            \fmfleft{i}
            \fmfright{o1}
            \fmf{wiggly}{i,m}
            \fmf{plain,s}{m,o1}
        \end{fmfgraph*}
        }
    &=
    \frac{1}{-i\omega + \nu q^2},&
    \parbox{15mm}{
        \begin{fmfgraph*}(15,10)
            \setval
            \fmfleft{i}
            \fmfright{o1}
            \fmf{plain}{i,m}
            \fmf{plain}{m,o1}
        \end{fmfgraph*}
        }
    &= \frac{2 D}{\omega^2 + \nu^2 q^4}, 
    \\
    \parbox{25mm}{
    \begin{fmfgraph*}(15,10)
        \setval
        \fmfleft{i}
        \fmfright{o1,o2}
        \fmf{wiggly}{i,m}
        \fmf{plain}{m,o1}
        \fmf{plain}{m,o2}
        \fmflabel{$\bm q_3\comma \omega_3$}{i}
        \fmfv{label=$\bm q_1\comma \omega_1$,label.angle=0}{o1}
        \fmfv{label=$\bm q_2\comma \omega_2$,label.angle=0}{o2}
    \end{fmfgraph*}
    }
    & =- \lambda \, \bm q_1 \cdot \bm q_2, &
    \parbox{25mm}{
    \begin{fmfgraph*}(15,10)
        \setval
        \fmfleft{i}
        \fmfright{o1,o2}
        \fmf{fermion}{i,m}
        \fmf{plain}{m,o1}
        \fmf{plain}{m,o2}
        \fmflabel{$\bm q_3\comma \omega_3$}{i}
        \fmfv{label=$\bm q_1\comma \omega_1$,label.angle=0}{o1}
        \fmfv{label=$\bm q_2\comma \omega_2$,label.angle=0}{o2}
        \fmfdot{m}
    \end{fmfgraph*}
    }    & = - i\omega_3 \beta \lambda \, \bm q_1 \cdot \bm q_2,
\end{align}
including the entropy-consumption vertex. The FDRs are the same as given in Eq.~\eqref{eq: GFDT matrix}. However, since there is only one field, we have $C = C^*$, and therefore the lower-right quadrant $\Delta_{22}$ is identically zero. Moreover, $G^\dagger = G^*$, which implies $\frac{i\omega}{D}\Delta_{21} =\Delta_{11} $.
Thanks to these simplifications, the only diagrams we need to calculate are as follows
\begin{align}
    \Sigma_{21} & \equiv
    \parbox{15mm}{
    \centering
    \begin{fmfgraph*}(15,10)
        \setval
        \fmfleft{i}
        \fmfright{o}
        \fmf{plain}{i,c1}
        \fmf{plain,left,tension=1/2}{c1,c2}
        \fmf{plain,left,tension=1/2}{c2,c1}
        \fmf{wiggly}{c2,o}
        \fmfdot{c1}
    \end{fmfgraph*}
    }
    =
    \parbox{15mm}{
    \centering
    \begin{fmfgraph*}(15,10)
        \setval
        \fmfleft{i}
        \fmfright{o}
        \fmf{fermion}{i,c1}
        \fmf{plain,left,tension=1/2}{c1,c2}
        \fmf{plain,left,tension=1/2}{c2,c1}
        \fmf{wiggly}{c2,o}
        \fmfdot{c1}
    \end{fmfgraph*}
    }
    +
    \parbox{15mm}{
    \centering
    \begin{fmfgraph*}(15,10)
        \setval
        \fmfleft{i}
        \fmfright{o}
        \fmf{plain}{i,c1}
        \fmf{fermion,right,tension=1/2}{c2,c1}
        \fmf{plain,left,tension=1/2}{c2,c1}
        \fmf{wiggly}{c2,o}
        \fmfdot{c1}
    \end{fmfgraph*}
    }
    +
    \parbox{15mm}{
    \centering
    \begin{fmfgraph*}(15,10)
        \setval
        \fmfleft{i}
        \fmfright{o}
        \fmf{plain}{i,c1}
        \fmf{plain,left,tension=1/2}{c1,c2}
        \fmf{fermion,left,tension=1/2}{c2,c1}
        \fmf{wiggly}{c2,o}
        \fmfdot{c1}
    \end{fmfgraph*}
    },
\end{align}
where the subscripts signal the external legs of the diagrams. Since these diagrams are amputated, we recover the full deviation functions by reattaching the legs. For example, $\Delta_{11}^* = G \Sigma_{22}^* G^* $.
The first integral is calculated as follows
\begin{align}
    \parbox{15mm}{
    \centering
    \begin{fmfgraph*}(15,10)
        \setval
        \fmfleft{i}
        \fmfright{o}
        \fmf{fermion}{i,c1}
        \fmf{plain,right,tension=1/2}{c2,c1}
        \fmf{plain,left,tension=1/2}{c2,c1}
        \fmf{wiggly}{c2,o}
        \fmfdot{c1}
    \end{fmfgraph*}
    }
    =
    \frac{i \omega  \lambda^2}{ D }
    \int_K (\bm k_1 \cdot \bm k_2)^2 C(\bm k_1, \omega_1)C(\bm k_2, \omega_2)
    = \frac{ 1 }{ 2 } i \omega \frac{D \lambda^2}{\nu^3} I_2 \equiv 2 i \omega \delta D, \quad\quad
    & I_n = \int_k \frac{ 1 }{ k^n }.
\end{align}
Here, $\delta D = 4 \frac{D \lambda^2}{\nu^3}I_2$ is exactly the renormalization of the noise-strength $D$ at one-loop level. The next two diagrams are equal. Furthermore, they have an external leg that includes a gradient, and consequently they contribute to a higher term as follows
\begin{align}
    \parbox{15mm}{
    \centering
    \begin{fmfgraph*}(15,10)
        \setval
        \fmfleft{i}
        \fmfright{o}
        \fmf{plain}{i,c1}
        \fmf{fermion,right,tension=1/2}{c2,c1}
        \fmf{plain,left,tension=1/2}{c2,c1}
        \fmf{wiggly}{c2,o}
        \fmfdot{c1}
    \end{fmfgraph*}
    }
    \sim \Oh(q^2).
\end{align}
Therefore, we obtain
\begin{align}
    \Sigma_{21} = 2 i \omega \delta D + \Oh(q^2)
    \implies
    \Sigma_{11} 
    = -\frac{ 1 }{ 2D }\left( G^*{}^{-1} \Sigma_{12} + G^{-1}\Sigma_{21} \right)
    = 2 \omega^2 \frac{ \delta D }{ D } .
\end{align}
Using Eq.~\eqref{eq: GFDT matrix}, and the identities we have found before, we obtain
\begin{align}
    \Delta_{12}^*
    = \frac{ D }{ i\omega } \Delta_{11}^*
    = \frac{ D }{ i\omega } G \Sigma_{11} G^*
    = - \frac{ 2 i\omega \delta D/D }{ \omega^2 + \nu^2 q^4 },
\end{align}
which verifies the FDR.

\section{Derivation of Eq. (24): Sum rules and generalized Harada-Sasa relations}

We can generalize the field-theoretic Harada-Sasa relation %~\cite{nardiniEntropyProductionField2017} 
to several fields and both non-conserved and conserved systems. If we consider a system as given in Eq.~(4) in the main text, then using the Onsager-Machlup functional we may write the susceptibility as
\begin{align}
    \chi_{ab}(\bm x -\bm x', t - t')
    \equiv
    \fdv{\E{\varphi_a(\bm x, t)}}{{h_b(\bm x', t')}}\bigg|_{h=0}
    =
    \frac{\beta}{2}
    \E{\varphi_a(\bm x, t)
    \left[ \partial_{t'}\varphi_b(\bm x', t') - \Gamma K_b[\varphi](\bm x', t') \right] }.
\end{align}
Taking the time- and space- anti-symmetric part, applying $\Gamma^{-1} \partial_t$, and using $G = \Gamma \chi$, we obtain
\begin{align}
    \partial_t
    \left[
    G_{ab}(\bm x, \bm t) - G_{ab}^\dagger(\bm x, \bm t) 
    - D^{-1}\partial_t C_{ab}(\bm x, \bm t)
    \right]
    =
    \frac{1}{2}
    \left[
    s_{ab}(\bm x, t)
    +
    s_{ab}^\dagger(\bm x, t)
    \right],
\end{align}
where we have defined the operator
\begin{align}
    s_{ab}(\bm x-\bm x', t - t')
    & = 
    \E{
    \beta \partial_t \varphi_a(\bm x, t) K_b[\varphi](\bm x', t')
    },
\end{align}
which is related to the total entropy $\E{S}$ by a trace operation,
\begin{align}
     \E{S}
    = \Tr\, s
    \equiv \int_{\bm x, \bm x', t, t'} 
    %\sum_{ab}
    \delta(\bm x - \bm x', t - t')
    \delta_{ab}\,
    s_{ab}(\bm x - \bm x', t - t')
    & = L^d T s_{aa}(\bm x = 0, t = 0)\\
    & = L^d T \int_{\bm q, \omega} s_{aa}(\bm q, \omega).
\end{align}
where $L$ is the linear dimension of the system and $T$ is the total (observation) time.
The Harada-Sasa relation can then be derived as follows
\begin{align}
    \Tr\, \partial_t
    \left[
    G_{ab}(\bm x, \bm t) - G_{ab}^\dagger(\bm x, t) 
    - D^{-1}\partial_t C_{ab}(\bm x, t)
    \right]
    =
    -
    \int_{\bm q, \omega} i \omega
    \left[
    G_{aa}(\bm q, \bm \omega) - G_{aa}^\dagger(\bm q, \omega) 
    - D^{-1}\partial_t C_{aa}(\bm x, \omega)
    \right]
    = \E{S},
\end{align}
where we have used the fact that $\Tr A^\dagger = \Tr A$ for real matrices.

We now apply the trace operator to the FDRs presented in Eq.~(1-3) in the main text, to obtain sum rules. First, using the fact that $C$ is Hermitian, we find
\begin{align}
    0 &= 
    \Tr \E{
    \varphi_a(\bm q, \omega) \varphi_b(\bm q', \omega')
    \left( e^{-S} - 1 \right)
    }.
\end{align}

Next, we obtain
\begin{align}
        \Tr
    \left[
    G_{ab}(\bm q, \omega) - G_{ab}^\dagger(\bm q, \bm \omega) 
    - \frac{i\omega}{D} C_{ab}(\bm q, \bm \omega)
    \right]
    & = 
    \Tr \E{
    \varphi_a(\bm q, \omega) i\tilde \varphi_b(\bm q', \omega')
    \left( e^{-S} - 1 \right)
    } ,\\
    \Tr\,\frac{i\omega}{D}
    \left[
    G_{ab}(\bm q, \omega) - G_{ab}^\dagger(\bm q, \bm \omega) 
    - \frac{i\omega}{D} C_{ab}(\bm q, \bm \omega)
    \right]
    & = 
    \Tr \E{
    i \tilde \varphi_a(\bm q, \omega) i\tilde \varphi_b(\bm q', \omega')
    \left( e^{-S} - 1 \right)
    }.
\end{align}
We observe that the last two identities bear a striking resemblance to the Harada-Sasa relation. This observation allows us to derive the following new identities
\begin{align}
   &\E{S}= - \Tr \,
    i\omega
    \E{
    \varphi_a(\bm q, \omega) i\tilde \varphi_b(\bm q', \omega')
    \left( e^{-S} - 1 \right)
    }, \\
    &\E{S}=-
    \Tr \,
    D
    \E{
    i \tilde \varphi_a(\bm q, \omega) i\tilde \varphi_b(\bm q', \omega')
    \left( e^{-S} - 1 \right)
    },
\end{align}
which should hold for any active field theory.

\end{fmffile}